\documentclass[journal=jacsat,manuscript=article]{achemso}
\setkeys{acs}{articletitle = true}
\setkeys{acs}{etalmode = truncate}
\setkeys{acs}{maxauthors = 10}
\usepackage{xcolor}
\usepackage{soul}

\usepackage[version=3]{mhchem} 

\author{Ulrich Pototschnig}
\affiliation{Department of Materials Science, Montanuniversität Leoben, Franz-Josef-Strasse 18, 8700 Leoben, Austria}
\alsoaffiliation{Department of Chemistry, University of Hamburg, Luruper Chaussee 149, 22761 Hamburg, Germany}
\email{ulrich.pototschnig@uni-hamburg.de}

\author{Martin Matas}
\affiliation{Department of Materials Science, Montanuniversität Leoben, Franz-Josef-Strasse 18, 8700 Leoben, Austria}

\author{David Scheiblehner}
\affiliation{Chair of Nonferrous Metallurgy, Montanuniversität Leoben, Franz-Josef-Strasse 18, 8700 Leoben, Austria}

\author{David Neuschitzer}
\affiliation{Chair of Nonferrous Metallurgy, Montanuniversität Leoben, Franz-Josef-Strasse 18, 8700 Leoben, Austria}

\author{Robert Obenaus-Emler}
\affiliation{Resources Innovation Center, Montanuniversität Leoben, Franz-Josef-Strasse 18, 8700 Leoben, Austria}

\author{Helmut Antrekowitsch}
\affiliation{Chair of Nonferrous Metallurgy, Montanuniversität Leoben, Franz-Josef-Strasse 18, 8700 Leoben, Austria}

\author{David Holec}
\affiliation{Department of Materials Science, Montanuniversität Leoben, Franz-Josef-Strasse 18, 8700 Leoben, Austria}
\email{david.holec@unileoben.ac.at}

\title{A Predictive Model for Catalytic Methane Pyrolysis}

\abbreviations{IR,NMR,UV}

\begin{document}

\begin{abstract}
Methane pyrolysis provides a scalable alternative to conventional hydrogen production methods, avoiding greenhouse gas emissions. However, high operating temperatures limit economic feasibility on an industrial scale. A major scientific goal is, therefore, to find a catalyst material that lowers operating temperatures, making methane pyrolysis economically viable. In this work, we derive a model that provides a qualitative comparison of possible catalyst materials. The model is based on calculations of adsorption energies using density functional theory. Thirty different elements were considered. Adsorption energies of intermediate molecules in the methane pyrolysis reaction correlate linearly with the adsorption energy of carbon. Moreover, the adsorption energy increases in magnitude with decreasing group number in the $d$-block of the periodic table. For a temperature range between 600 and 1200 K and a normalized partial pressure range for $\mathrm{H_2}$ between $10^{-1}$ and $10^{-5}$, a total of eighteen different materials were found to be optimal catalysts at least once. This indicates that catalyst selection and reactor operating conditions should be well-matched. The present work establishes the foundation for future large-scale studies of surfaces, alloy compositions, and material classes using machine learning algorithms.
\end{abstract}

\section{Introduction}
Hydrogen is an important energy carrier and chemical feedstock and has the potential to replace fossil hydrocarbons in industry and transportation. However, hydrogen production based on fossil fuels accounted for almost the entire production in 2021, with associated emissions of over 900 Mt of \ce{CO_2}. Although emission-free hydrogen production from water electrolysis has accelerated over the past few years, it covers only 0.1\% of global demand \cite{iea2023}. Thus, an economically and ecologically feasible bridging technology for hydrogen production is needed.

As suggested by \citeauthor{SanchezBastardo2021}, a promising option is methane pyrolysis, i.e., thermal decomposition of methane in the absence of oxygen to form solid carbon and hydrogen \cite{SanchezBastardo2021}.
\begin{equation}\label{methane_pyrolysis}
    \mathrm{CH_{4} \, (g)} \rightarrow \mathrm{2 \, H_{2} \, (g)} + \mathrm{C \, (s)}
\end{equation}
While methane pyrolysis still relies on the extraction of natural gas, the process is \ce{CO_2} emission-free. From an energetic point of view, only 37.5 kJ of energy is necessary to produce 1 mol of \ce{H_2} via methane pyrolysis, compared to 286 kJ required for water electrolysis \cite{Blanksby2003}. Even though a significant reaction yield can be achieved only above 1200 °C \cite{Ashik2015}, this figure can be substantially reduced by using a catalyst. There is a multitude of research efforts to develop an efficient and cost-effective catalyst \cite{McConnachie2023}, including metal- and carbon-based solids \cite{Fan2021, Abbas2010, Suelves2008}, molten salts \cite{Kang2020, Parkinson2021} as well as molten metals \cite{Zeng2020} and alloys \cite{Upham2017, Palmer2019, Sorcar2023, Scheiblehner2023_1, Scheiblehner2023_2}. In particular, pyrolysis in liquid bubble column reactors holds a big promise for sustainable process which through its design circumvents catalyst deactivation as the carbon that accumulates on top of the metal bath can be continuously removed from the reactor.

Ab initio methods provide a useful tool for high-throughput screening of material choices. In this work, a straightforward and flexible model for qualitative comparison of possible catalysts is derived. The reaction will be divided into its elementary steps, and a combination of the microkinetic model and Sabatier analysis will be used to describe the reaction rate as a single function of temperature $T$, pressure $p$ and adsorption energy of carbon $\Delta E_\mathrm{Ads, C}$ which is applied to thirty different elements.

\section{Model}
\subsection{Methane pyrolysis}
The methane pyrolysis reaction and its individual reaction steps have been widely studied in the past \cite{Ashik2015, Snoeck1997}. A catalytic decomposition process follows seven different elementary reaction steps, shown below, with the last step occurring twice per molecule. The sign `$^{*}$' denotes a catalyst surface site, e.g., $\mathrm{CH_{4}^{*}}$ is a methane molecule adsorbed on a catalyst. Methane chemisorbs on the catalyst surface.
\begin{equation}\label{CH4_adsorption}
    \mathrm{CH_{4} \; (g)} + ^{*} \rightarrow \mathrm{CH_{4}^{*}}
\end{equation}
Chemisorbed methane dissociates into a methyl radical ($\mathrm{CH_{3}}$) and a hydrogen atom.
\begin{equation}\label{diss_step_0_old}
    \mathrm{CH_{4}^{*}} + ^{*} \rightarrow \mathrm{CH_{3}^{*}} + \mathrm{H^{*}}
\end{equation}
The methyl radical dissociates further into a methylene ($\mathrm{CH_{2}}$), and then into a methine ($\mathrm{CH}$) radical, followed by the last dissociation step to form adsorbed elemental carbon and hydrogen.
\begin{equation}\label{diss_step_1}
    \mathrm{CH_{3}^{*}} + ^{*} \rightarrow \mathrm{CH_{2}^{*}} + \mathrm{H^{*}}
\end{equation}
\begin{equation}\label{diss_step_2}
    \mathrm{CH_{2}^{*}} + ^{*} \rightarrow \mathrm{CH^{*}} + \mathrm{H^{*}}
\end{equation}
\begin{equation}\label{diss_step_3}
    \mathrm{CH^{*}} + ^{*} \rightarrow \mathrm{C^{*}} + \mathrm{H^{*}}
\end{equation}
While carbon desorbs to form solid carbon, the hydrogen atoms recombine to produce hydrogen gas.
\begin{equation}\label{C_desorption}
    \mathrm{C^{*}} \rightarrow \; \mathrm{C\;(s)}+ \; ^{*} 
\end{equation}
\begin{equation}\label{H_desorption}
    2 \mathrm{H^{*}} \rightarrow \mathrm{H_{2}\;(g)}+ \; 2 ^{*} 
\end{equation}

While this reaction path is generally accepted, it is still disputed which reaction step is the slowest and, therefore, rate-determining for the overall reaction. In the 1960s, \citeauthor{Kozlov1962} concluded that Eq.~(\ref{diss_step_1}) is the rate-limiting step \cite{Kozlov1962}, whereas in \citeyear{BAKER1972}, \citeauthor{BAKER1972} considered it to be carbon diffusion \cite{BAKER1972}. More recently, surface transport phenomena are also suggested \cite{Hofmann2005}.
Computational studies using Density Functional Theory calculations found Eq. (3) to be rate-determining Fan et al., which will be used in this work.Computational studies using Density Functional Theory calculations carried out by \citeauthor{Liao1998} and \citeauthor{Fan2012} both demonstrated that Eq.~(\ref{diss_step_3}) exhibits the highest activation energies across various systems, encompassing both mono- and bimetallic surfaces \cite{Liao1998, Fan2012}. Therefore, conducting further studies on this aspect would be redundant. Despite the inherent limitations of the model, the assumption of a rate-limiting step based on the highest activation energy is reasonable, especially for high activation barriers.

Two simplifications are introduced. Firstly, as shown by \citeauthor{Fan2012}, Eqs.~(\ref{CH4_adsorption}) and (\ref{diss_step_0_old}) are combined since the energy barrier for initial adsorption of methane is negligibly small \cite{Fan2012}. Hence, the first reaction step is
\begin{equation}\label{diss_step_0}
    \mathrm{CH_{4}\;(g)}  +  \mathrm{2^{*}}  \rightarrow  \mathrm{CH_{3}^{*}}  +  \mathrm{H^{*}},
\end{equation}
reducing the number of steps to six. Secondly, the role of carbon desorption will be neglected. Carbon deposition on a catalyst surface is a complex phenomenon that can induce the formation of a wide variety of structures, such as carbon nanotubes, among many others \cite{Helveg2004, Chen2005}. Thus, for practical reasons, the role of carbon in this context is greatly simplified by omitting Eq.~(\ref{C_desorption}) and setting the activity of carbon $a_\mathrm{C}$ to 1.

\subsection{Sabatier analysis}\label{sec:sabatier}
From the Sabatier principle, it follows that stronger adsorption of a reactant, i.e., a higher magnitude of the negative adsorption energies $\Delta E_\mathrm{Ads}$, correlates with a higher reaction rate constant $\mathrm{k}$. This is valid up to a point where too strong adsorption inhibits product desorption, resulting in a lack of free surface sites $\theta_*$ \cite{Knoezinger2003}. Together with the Br\o{}nsted relationship that links thermodynamics and kinetics of a reaction \cite{Bronsted1924}, both concurring mechanisms caused by adsorption are responsible for the characteristic volcano-shaped plots of catalytic reaction rates that were first reported in \citeyear{Balandin1969} \cite{Balandin1969}. In addition, the findings of \citeauthor{AbildPedersen2007} \cite{AbildPedersen2007} showing that the adsorption energy of \ce{CH_$x$} ($x = 1, 2, 3$) scales approximately with the adsorption energy of a C atom are applied to simplify the model, resulting in the reaction rate as a function of only the adsorption energy of carbon $\Delta E_{\mathrm{Ads,\,C}}$.

As a result, the system of elementary steps that was established above contains five reaction steps, with the fourth step (Eq.~(\ref{diss_step_3})) being rate-determining based on findings by \citeauthor{Fan2012} \cite{Fan2012}. Four equilibrium constants can be derived:
\begin{equation}\label{K_1}
    \mathrm{from\; Eq.~(\ref{diss_step_0})}: \mathrm{K_1} = \cfrac{\theta_\mathrm{CH_{3}} \theta_\mathrm{H}}{p_\mathrm{CH_{4}}\theta_{*}^2},
\end{equation}
\begin{equation}\label{K_2}
     \mathrm{from\; Eq.~(\ref{diss_step_1})}: \mathrm{K_2} = \cfrac{\theta_\mathrm{CH_{2}} \theta_\mathrm{H}}{\theta_\mathrm{CH_{3}}\theta_{*}},
\end{equation}
\begin{equation}\label{K_3}
     \mathrm{from\; Eq.~(\ref{diss_step_2})}: \mathrm{K_3} = \cfrac{\theta_\mathrm{CH} \theta_\mathrm{H}}{\theta_\mathrm{CH_{2}}\theta_{*}},
\end{equation}
\begin{equation}\label{K_5}
     \mathrm{from\; Eq.~(\ref{H_desorption})}: \mathrm{K_5} = \cfrac{p_\mathrm{H_{2}} \theta_{*}^2}{\theta_\mathrm{H}^2}
\end{equation}
where $\mathrm{K}_i$ is the equilibrium constant of reaction $i$, $\theta_{j}$ is the surface coverage of species $j$, $p_\mathrm{CH_{4}}$ and $p_\mathrm{H_{2}}$ are partial pressures of $\mathrm{CH_{4}}$ and $\mathrm{H_{2}}$, respectively. Partial pressures are normalized with respect to the total system pressure. $\mathrm{K_{4}}$ is missing because it is in a non-equilibrium state. Equilibrium constant and Gibbs free energy are connected via
\begin{equation}\label{eq:Ki}
    \mathrm{K}_{i} = \mathrm{exp}\left(-\cfrac{\Delta G_{i}^{\circ}}{k_\mathrm{B} T}\right)
    = \mathrm{exp}\left(-\cfrac{\Delta E_{i}^{\circ}}{k_\mathrm{B} T}+\cfrac{\Delta S_{i}^{\circ}}{k_\mathrm{B}} \right).
\end{equation}
For any intermediate steps where no gas-phase molecules are being adsorbed or desorbed, namely (\ref{K_2}) and (\ref{K_3}), the entropy contributions are ignored as the changes in entropy are negligible, i.e.,
\begin{equation}\label{Eq:G-approx-E}
    \Delta G_i  \approx \Delta E_i.
\end{equation}
Reaction steps (\ref{K_1}) and (\ref{K_5}) require a different approach. As shown by \citeauthor{Norskov2014}, to a good approximation, adsorbed molecules can be expected to lose all their entropy upon adsorption \cite{Norskov2014}:
\begin{equation}\label{entropy-approx}
    \Delta S_{\mathrm{Ads}}^\circ = S_{\mathrm{Ads}}^\circ - S_{\mathrm{gas}}^\circ \approx - S_{\mathrm{gas}}^\circ.
\end{equation}
Consequently, the same must apply in reverse for desorption.

For Eq.~(\ref{K_1}) to Eq.~(\ref{K_5}) there are now five unknown variables, i.e. all coverages $\theta_{j}$, with only four equations. Rearranging the system of equations above yields four coverages, which are all functions of $\theta_{*}$. 
\begin{equation}\label{theta_H}
    \text{From Eq. (\ref{K_5}):} \quad \theta_\mathrm{H} = \sqrt{\cfrac{p_\mathrm{H_2}}{\mathrm{K_{5}}}} \theta_{*}
\end{equation}
\begin{equation}\label{theta_CH3}
    \text{Eq. (\ref{K_1}) with Eq. (\ref{theta_H}):} \quad \theta_\mathrm{CH_{3}} = \mathrm{K_1} p_\mathrm{CH_4} \sqrt{\cfrac{\mathrm{K_{5}}}{p_\mathrm{H_2}}} \theta_{*}
 \end{equation}
\begin{equation}\label{theta_CH2}
    \text{Eq. (\ref{K_2}) with Eq. (\ref{theta_CH3}):} \quad \theta_\mathrm{CH_{2}} = \mathrm{K_1 \; K_2 \; K_5} \cfrac{p_\mathrm{CH_4}}{p_\mathrm{H_2}} \theta_{*}
\end{equation}
\begin{equation}\label{theta_CH}
    \text{and Eq. (\ref{K_3}) with Eq. (\ref{theta_CH2}):} \quad \theta_\mathrm{CH} = \mathrm{K_1 \; K_2 \; K_3 \; K_5^{\frac{3}{2}}} \cfrac{p_\mathrm{CH_4}}{p^{\frac{3}{2}}_\mathrm{H_2}} \theta_{*}.
\end{equation}
The site conservation rule yields a fifth equation with its general form of
\begin{equation}
    \theta_{*} + \sum_{j \neq *}\theta_{j} = 1,
\end{equation}
where a fraction of $j$-covered sites can be defined as $\lambda_{j} = \theta_{j}/\theta_{*}$. This leads to
\begin{equation}
    \theta_{*} \left(1 + \sum_{j \neq *}\lambda_{j} \right) = 1
\end{equation}
and the coverage of free sites can then be expressed as
\begin{equation}
    \theta_{*} =  \left(1 + \sum_{j \neq *}\lambda_{j} \right)^{-1}.
\end{equation}
Inserting all terms from above yields the coverage of free surface sites for the system as
\begin{equation}\label{theta_star}
    \theta_{*} = \left(1 +
    \mathrm{K_1} p_\mathrm{CH_4} \sqrt{\cfrac{\mathrm{K_{5}}}{p_\mathrm{H_2}}} +
    \mathrm{K_1 K_2 K_5}  \cfrac{p_\mathrm{CH_4}}{p_\mathrm{H_2}} + \mathrm{K_1 K_2 K_3 K_5^{\frac{3}{2}}} \cfrac{p_\mathrm{CH_4}}{p^{\frac{3}{2}}_\mathrm{H_2}} +
    \sqrt{\cfrac{p_\mathrm{H_2}}{\mathrm{K_{5}}}}\right)^{-1}.
\end{equation}
Now, the only non-equilibrated reaction step (Eq.~(\ref{diss_step_3})) is being considered. For this, a reaction rate expression can be set up as
\begin{equation}\label{R_4}
    R_{4} = \mathrm{k_{4}}  \theta_\mathrm{CH} \theta_{*}  -  \mathrm{k_{-4}}  \theta_\mathrm{H} a_\mathrm{C}
\end{equation}
with $a_\mathrm{C} = 1$. It is not possible to define an equilibrium constant for Eq.~(\ref{R_4}). However, it is convenient to define an ``approach to equilibrium'' $\mathrm{\gamma}$ \cite{Norskov2014}. This approach to equilibrium is a positive number that provides information about whether the reaction proceeds in a forward or backward direction.
\begin{flalign*}
       & \mathrm{\gamma} < 1:\quad \text{The reaction proceeds  in  the forward direction.}         \\
       & \mathrm{\gamma} = 1:\quad \text{The reaction is in equilibrium.}        \\
       & \mathrm{\gamma} > 1:\quad \text{The reaction proceeds in the backward direction.} 
\end{flalign*}
Hence, the approach to equilibrium $\mathrm{\gamma_{4}}$ is defined as
\begin{equation}
    \mathrm{\gamma_{4}} \mathrm{K_{4}} = \mathrm{\gamma_{4}}\cfrac{\mathrm{k_4}}{\mathrm{k_{-4}}} = \cfrac{\theta_\mathrm{H} a_\mathrm{C}}{\theta_\mathrm{CH} \theta_{*}}
\end{equation}
leading to a simplified reaction rate of
\begin{equation}\label{R_4_final}
    R_{4} = \mathrm{k_{4}} \mathrm{\theta_{CH}} \mathrm{\theta_{*}} (1 -  \mathrm{\gamma_{4}})
\end{equation}
with $\mathrm{k_{4}}$ defined as
\begin{equation}\label{k_4}
    \mathrm{k_{4}} = \cfrac{k_\mathrm{B} T}{\mathrm{h}} \; \mathrm{exp}\left(-\cfrac{\Delta E_{4}^\mathrm{A}}{k_\mathrm{B} T} +\cfrac{\Delta S^\mathrm{A}_4}{k_\mathrm{B}}\right).
\end{equation}
Having only one rate-determining reaction step allows for the important simplification of
\begin{equation}
    \mathrm{\gamma_{4}} \approx \gamma \rightarrow R_{4} \approx R
\end{equation}
where $\gamma$ is the approach to equilibrium for the overall reaction rate, i.e.,
\begin{equation}
    \gamma = \cfrac{p_\mathrm{H_2}^2}{{\mathrm{K_{eq}} p_\mathrm{CH_4}}} \quad \text{with} \quad \mathrm{K_{eq}} = \mathrm{exp}\left(-\cfrac{\Delta G}{k_\mathrm{B} T} \right),
\end{equation}
and $p_\mathrm{CH_4}$ and $p_\mathrm{H_2}$ are normalized partial pressures of $\mathrm{CH_4}$ and $\mathrm{H_2}$, respectively, $\mathrm{K_{eq}}$ is the equilibrium constant, and $\Delta G$ is the Gibbs free energy of the overall reaction. Using these simplifications in Eq.~(\ref{R_4_final}) leads to the reaction rate as
\begin{equation}
    R(T, p, \Delta E_\mathrm{Ads, C})= \cfrac{k_\mathrm{B} T}{\mathrm{h}}
    \exp\left(\cfrac{\Delta S^\mathrm{A}_4}{k_\mathrm{B}}\right) \exp \left(-\cfrac{\Delta E^\mathrm{A}_4}{k_\mathrm{B} T}\right) \theta_\mathrm{CH} \theta_{*} (1-\gamma)
\end{equation}
with $\Delta S^\mathrm{A}_4$ and $\Delta E^\mathrm{A}_4$ as the reaction entropy and energy of non-equilibrated reaction step 4. While
$\Delta E^\mathrm{A}_4$ follows the transition-state scaling relation that is derived from adsorption energy trends \cite{AbildPedersen2007}, $\Delta S^\mathrm{A}_4$ is neglected for step 4 (Eq.~(\ref{diss_step_3})), i.e. $\Delta S^\mathrm{A}_4 = 0$ because reactants and products stay in an adsorbed state. This leads to the final reaction rate equation
\begin{equation}\label{reaction_rate}
    R(T, p, \Delta E_\mathrm{Ads, C}) = \cfrac{k_\mathrm{B} T}{\mathrm{h}}
    \exp\left(-\cfrac{\Delta E^\mathrm{A}_{4}}{k_\mathrm{B} T}\right) \theta_\mathrm{CH} \theta_{*} (1 - \gamma).
\end{equation}
Impressively, this model describes the reaction rate as a combination of microscopic quantities, such as adsorption and activation energy, and macroscopic properties, such as temperature and pressure. Importantly, the microscopic quantities can be obtained using ab initio calculations. From $R$, the volcano-shaped curve is obtained, the maximum position of which corresponds to adsorption energy, which in turn is characteristic of a particular ideal catalyst material.

\section{Methodology}
Adsorption, reaction, and activation energies for catalytic methane pyrolysis were calculated using density functional theory (DFT) calculations \cite{Hohenberg1964}. Values for standard enthalpy and entropy of the total reaction were taken from the National Institute of Standards and Technology (NIST) \cite{NIST}. The Vienna Ab initio Simulation Package (VASP) was used for all DFT calculations in this work \cite{Kresse1993-xm}. 
Exchange-correlation functional parametrized by Perdew, Burke, and Ernzerhof (PBE-GGA) \cite{Perdew1996} and revised by \citeauthor{Hammer1999}\cite{Hammer1999} to improve chemisorption energetics of atoms and molecules on transition-metal surfaces was employed. A projector augmented wave (PAW) method was used for electron-ion interactions \cite{Bloechl1994}. Magnetic moments for ferro- or paramagnetic substrates were taken from the Materials Project (www.materialsproject.org), an open dataset for properties of inorganic materials \cite{Jain2013}. A supercell of $3\times 3\times 4$ ($2\times2\times4$) primitive cells was used to generate the cubic (hexagonal) slabs for the surface calculations, thus leading to (111), (110), and (0001) surfaces for fcc, bcc, and hcp metals, respectively. These represent the closely packed plane for the considered crystal systems. The additional vacuum of more than 20 \r{A} was inserted to separate the periodic images of the free surfaces. $\Gamma$-centered $k$-mesh of $6\times 6\times 1$ was chosen. Default plane wave cutoff energy from the PAW pseudopotentials\cite{Kresse1999-if} was used. More detailed information on the calculations carried out can be found in the supporting material.

\section{Results}
\subsection{Adsorption energies}
Calculated adsorption energies are presented below. Fig. \ref{cp-aet} depicts adsorption energies of CH, \ce{CH2}, \ce{CH3}, and H, respectively, as a function of carbon adsorption energy for a given substrate element. In the lower right corners of each graph, the Pearson correlation coefficient and the linear fit function, which will be used for the Sabatier model later, are shown. The three crystal structures considered, face-centered cubic (fcc), body-centered cubic (bcc), and hexagonal close-packed (hcp), have all close-packed surface structures, i.e., (111) for fcc, (110) for bcc and (0001) for hcp. The results do not show any clear difference between the behavior depending on the crystallography of the substrate: the adsorption energies are evenly spread across the whole energy range, with a possible exception of the bcc systems more accumulated in the low energy regions. The linear correlations are acceptable in all four cases, i.e., the dependence of the adsorption energy of CH, \ce{CH2}, \ce{CH3}, and H on the adsorption energy of C. The best linear correlation is obtained between $\Delta E_{\mathrm{Ads, CH_2}}$ and $\Delta E_{\mathrm{Ads, C}}$ ($R=0.93$), whereas the weakest correlation is obtained between $\Delta E_{\mathrm{Ads, H}}$ and $\Delta E_{\mathrm{Ads, C}}$ ($R=0.81$).

\begin{figure}[H]
  \centering
  \includegraphics[width=0.98\textwidth]{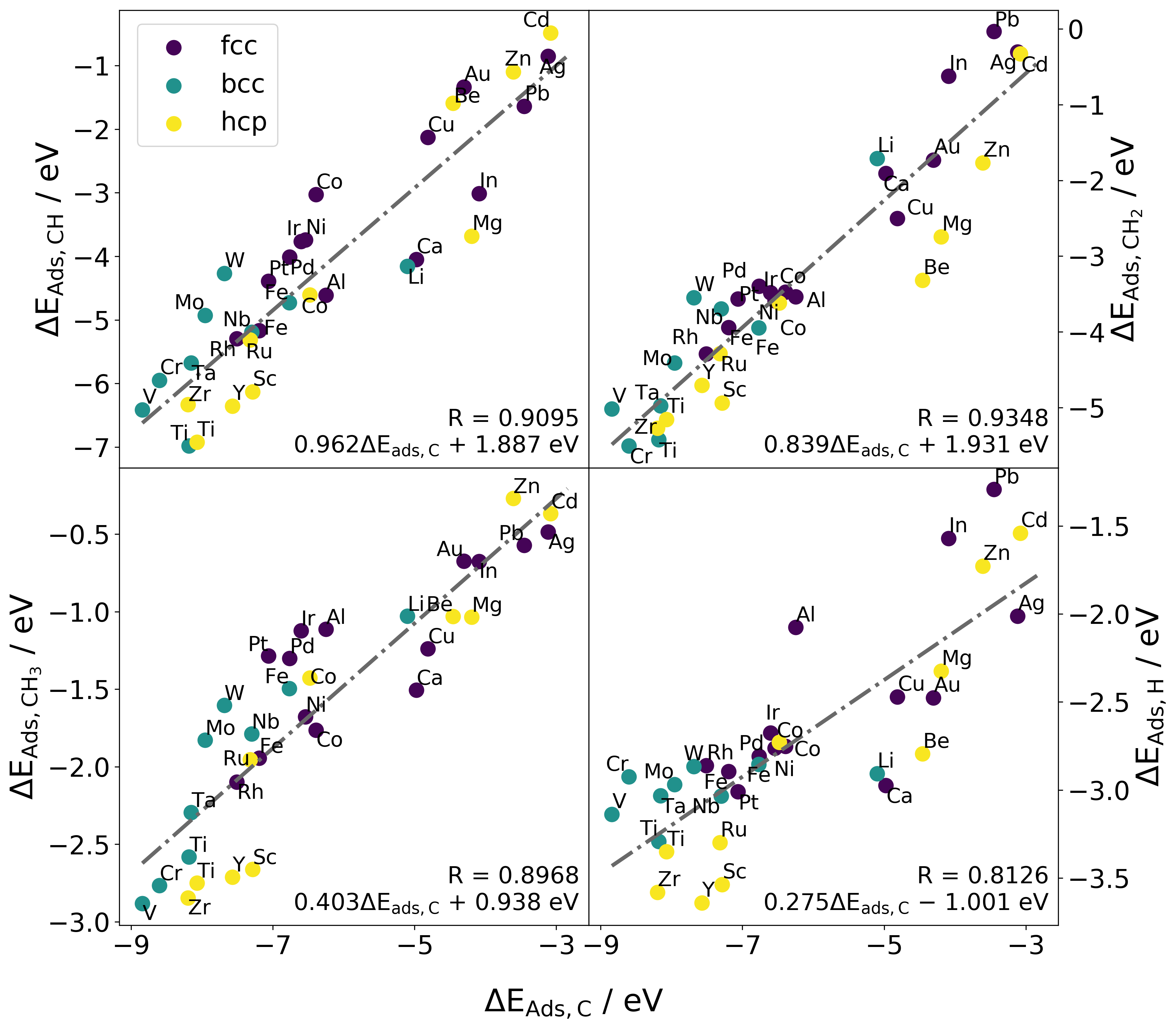}
  \captionsetup{width=.95\linewidth, justification=centering}
  \caption{Adsorption energies of CH, \ce{CH2}, \ce{CH3}, and H as a function of $\Delta E_\mathrm{Ads, C}$ for fcc, bcc, and hcp substrate structures with close-packed surface planes.}
  \label{cp-aet}
\end{figure}

\begin{figure}[H]
  \centering
  \includegraphics[width=0.98\textwidth]{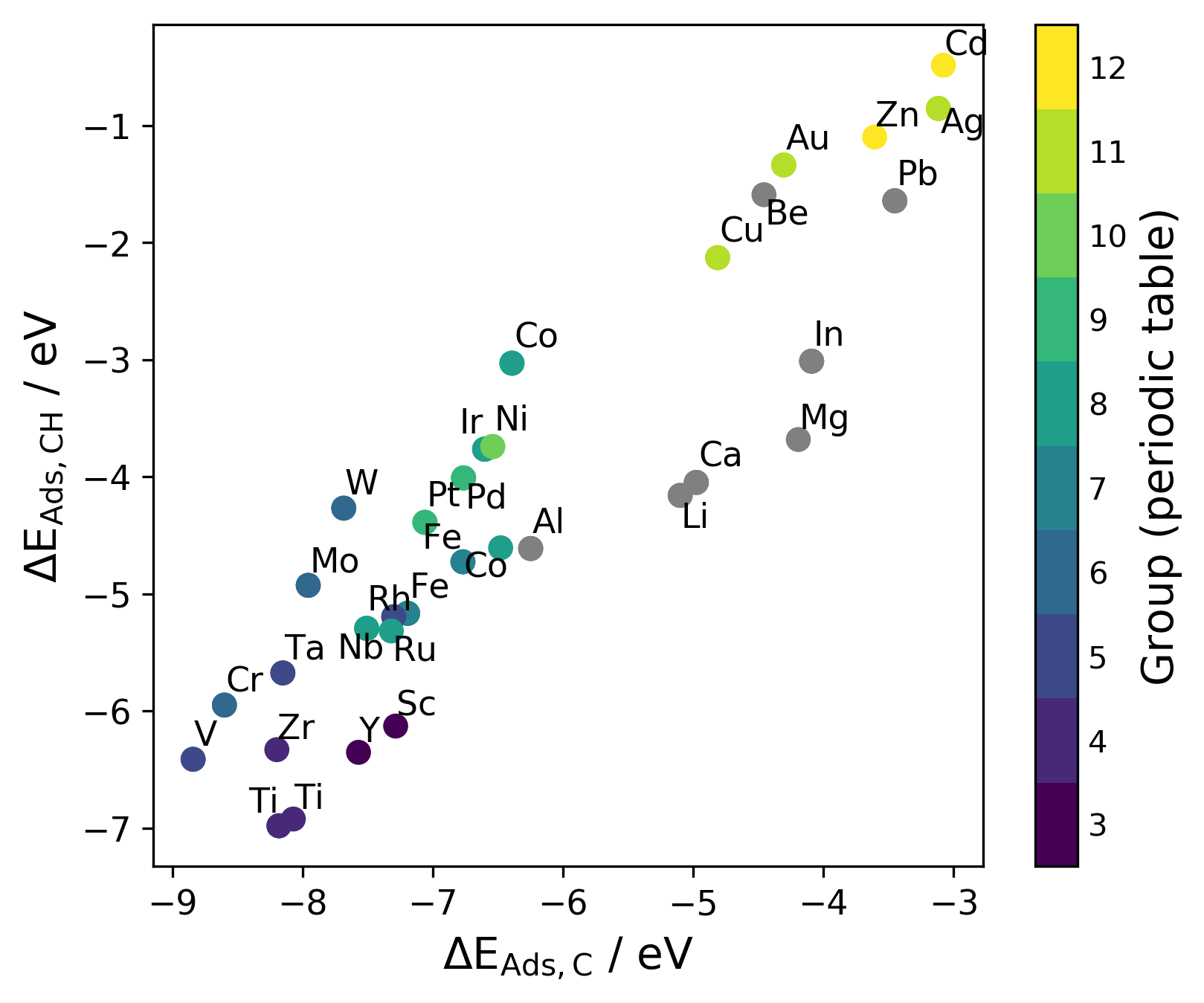}
  \captionsetup{width=.95\linewidth, justification=centering}
  \caption{$\Delta E_\mathrm{Ads, CH}$ as a function of $\Delta E_\mathrm{Ads, C}$ for different substrates. For $d$-block elements, the correlation between group number and their adsorption energy is clearly evident.}
  \label{d-block}
\end{figure}

A striking correlation between adsorption energies and position in the $d$-block of the periodic table can be seen in Fig. \ref{d-block}, where each element is color-coded according to its group on the example of the $\Delta E_{\mathrm{Ads, CH}}$ vs. $\Delta E_{\mathrm{Ads, C}}$ presented in Fig. \ref{cp-aet}. Elements that are not included in the $d$-block are grayed out. One explanation could be due to the fact that transition metals are mainly characterized by their partially filled $d$-subshells. Due to the directional character of $d$-orbitals, the nucleus is weakly shielded, interactions between $d$-electrons are weak, and the nucleus does not only strongly attract $d$- but also $s$-electrons from the next higher $s$-orbital. This results in relatively high but slowly increasing ionization energies over a given period \cite{Siekierski2002}, which in turn inhibits reactivity. This is evident when considering the enthalpy of hydration, which also decreases in magnitude with a higher group number \cite{Deeth2010}.

\subsection{Activation energies}
\citeauthor{Norskov2014}\cite{Norskov2014} proposed that the adsorption and activation energies should be correlated since the same fundamental physics governs them. \citeauthor{AbildPedersen2007}\cite{AbildPedersen2007} showed that in the case of methane, this correlation is exclusively related to the $\Delta E_{\mathrm{Ads, C}}$ term. The activation energies are needed as inputs for evaluating the reaction rate $R$ (Eq.~\eqref{reaction_rate}). Namely, $E^{\mathrm{A}}_4$ corresponding to the dissociation of the CH molecule, the rate-determining step, Eq.~\eqref{diss_step_3}, is needed. The explicit activation energy calculations were carried out for a (111)-surface of fcc-Cu. Details can be found in the supporting material. The resulting relationship is
\begin{equation}
    E_4^{\mathrm{A}} = 0.268\Delta E_{\mathrm{Ads, C}} + 4.7\,\text{eV}\ .
\end{equation}

\subsection{Reaction energies}
The remaining ingredient in the free site coverage, $\theta_*$, depends on the equilibrium constants K$_1$, K$_2$, K$_3$, and K$_5$. For their evaluation through Eq.~\eqref{eq:Ki}, the Gibbs free energy of a reaction is needed. As discussed in Section ``Sabatier analysis'', different treatment is applied depending on whether the reactions involve gas-phase molecules. 

Similar to the activation energies, reaction energies can also be correlated with adsorption energies. The reaction calculations were again carried out for a (111) surface of fcc-Cu and are detailed in the supporting material. The resulting relationships (Eq.~\eqref{Eq:G-approx-E}) are
\begin{equation}
    \Delta G_2\approx\Delta E_2 = 0.436\Delta E_{\mathrm{Ads, C}} + 4.28\,\text{eV}\ ,
\end{equation}
and
\begin{equation}
    \Delta G_3\approx\Delta E_3 = 0.123\Delta E_{\mathrm{Ads, C}} + 2.57\,\text{eV}\ .
\end{equation}

Reaction energies for $\ce{CH4}(\text{g})\to\ce{CH3}(\text{g})+\ce{H}(\text{g})$ and $2\ce{H}(\text{g})\to\ce{H2}(\text{g})$ were calculated as 4.725~eV and $-4.479$~eV, respectively. Using standard entropy values for the term $-S^\circ_{\mathrm{gas}}$ from \citeauthor{Lide2004-ev}\cite{Lide2004-ev}, the remaining Gibbs free energies of reaction are
\begin{equation}
    \Delta G_1 = 4.725\,\text{eV} + T\times0.00193\,\text{eV/K}\ ,
\end{equation}
and
\begin{equation}
    \Delta G_5 = -4.479\,\text{eV} - T\times0.00135\,\text{eV/K}\ .
\end{equation}

\subsection{Reaction rate}
\begin{figure}[H]
    \centering
    \includegraphics[width=0.96\textwidth]{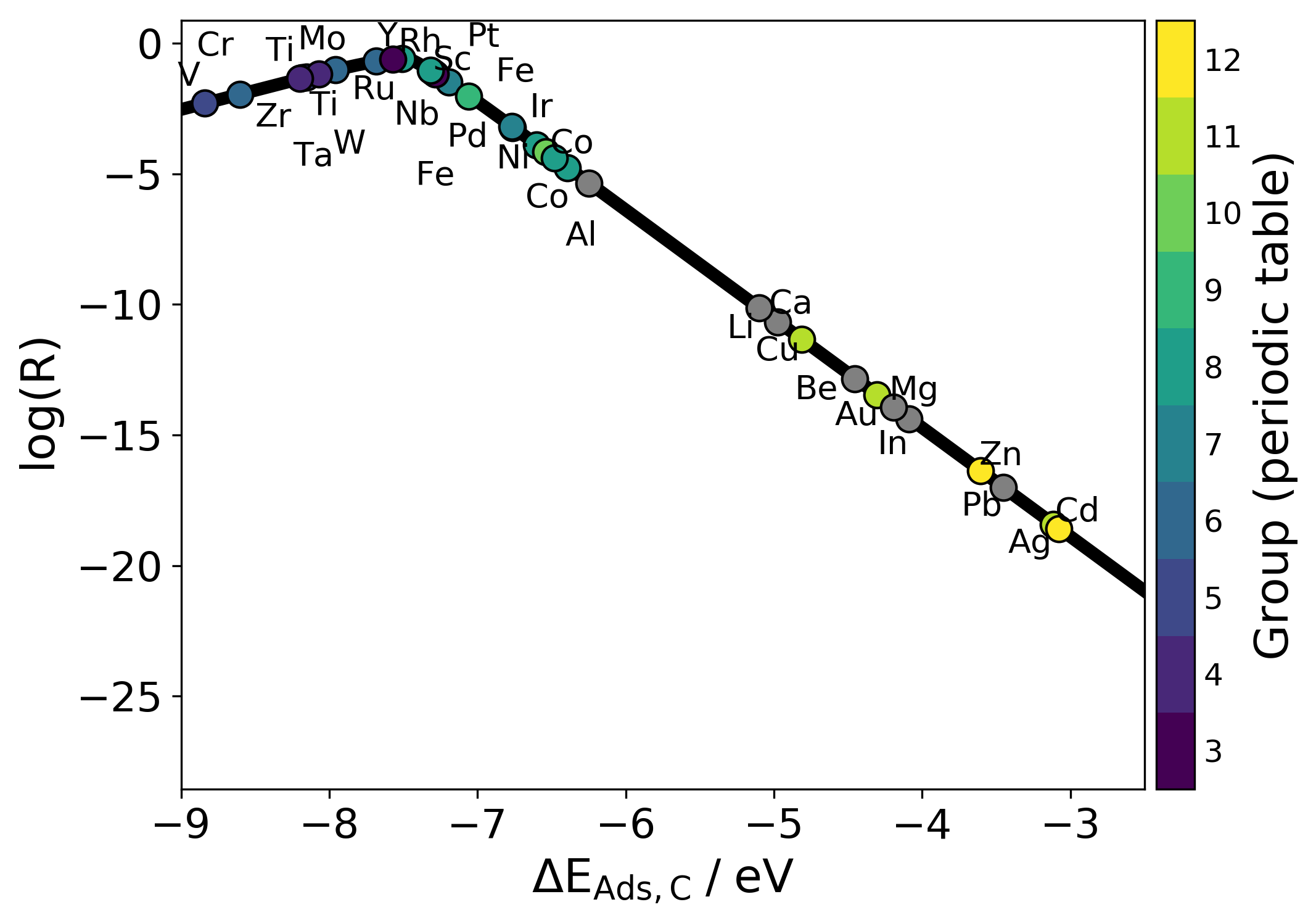}
    \captionsetup{width=.8\linewidth, justification=centering}
    \caption{Logarithm of the reaction rate, $R$, as a function of $\Delta E_\mathrm{Ads, C}$ for $T = 1000$ K, $p_{\ce{CH4}} = 0.99$, and $p_{\ce{H2}} = 0.01$.}
    \vspace{0pt}
    \label{R_first}
\end{figure}

The reaction rate, $R$, is now calculated from the adsorption energy trends, reaction, and activation energies. Since $R= R(T, p, \Delta E_\mathrm{Ads, C})$ (cf. Eq.~\eqref{reaction_rate}), representative thermodynamic input parameters are chosen: $T = 1000$ K, $p_{\ce{CH4}} = 0.99$, and $p_{\ce{H2}} = 0.01$. The resulting reaction rate is shown in Fig. \ref{R_first} as a function of $\Delta E_{\mathrm{Ads, C}}$, which can be linked with various base metals via the data presented in Fig.~\ref{cp-aet}. 

The best-performing surfaces are Rh, along with Ru, Y, Sc, Nb, and W, which are also close to the maximum. Overall, the occurrence of materials around the maximum is very high, which implies a wide range of options for close-to-optimum performance under those conditions. Furthermore, the apparent ``reaction gap'' around $\Delta E_{\mathrm{Ads, C}}=-6$ eV may serve as a practical cutoff for material selection.

\section{Discussion}
\begin{figure}[H]
    \vspace{-10pt}
    \captionsetup{width=.9\linewidth, justification=centering}
    \centering
    \includegraphics[width=0.95\textwidth]{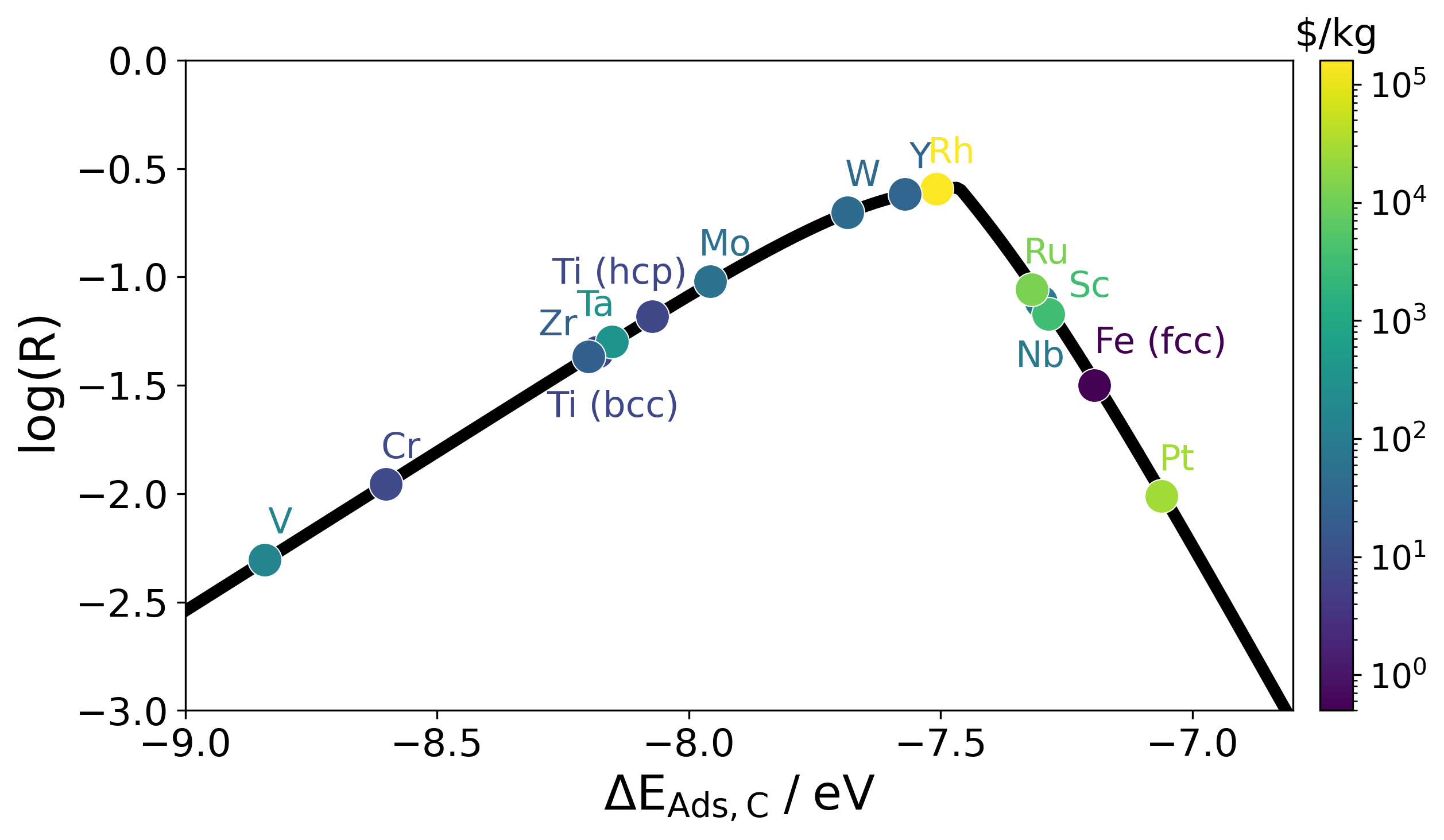}
    \caption{Region around the maximum log($R$) for $T = 1000$ K, $p_{\ce{CH4}} = 0.99$, and $p_{\ce{H2}} = 0.01$ with color-coded markers according to the global market price.}
    \label{cat_prices}
\end{figure}

Figure \ref{cat_prices} shows the elements exhibiting the highest reaction rate according to the Fig.~\ref{R_first} and color-coded according to their global market price as of February 2024 \cite{Metals} (in US Dollar (\$) per kg). For economic upscaling of catalytic methane pyrolysis to industrial application, a price consideration will be of great importance. Hence, price-informed volcano plots offer a straightforward way to make strategic decisions for a given set of calculated catalyst surfaces with user-defined operating conditions.

Figure \ref{R(T, p)} shows $R$ as a function of different temperatures and pressures. The upper plot demonstrates that the slope of $\log(R)$ (as a function of $\Delta E_{\mathrm{Ads, C}}$) decreases with increasing temperature while the maximum is shifted to lower magnitudes of adsorption energy. Overall, the rate increases with temperature for otherwise the same conditions, as intuitively expected. However, we note that increasing temperature to increase the yield has its practical and economical limits. The lower graph shows that the pressure also serves as a tool to tune the maximum reaction rate (optimal operating conditions) to other elements.

\begin{figure}[H]
    \centering
    \includegraphics[width=0.9\textwidth]{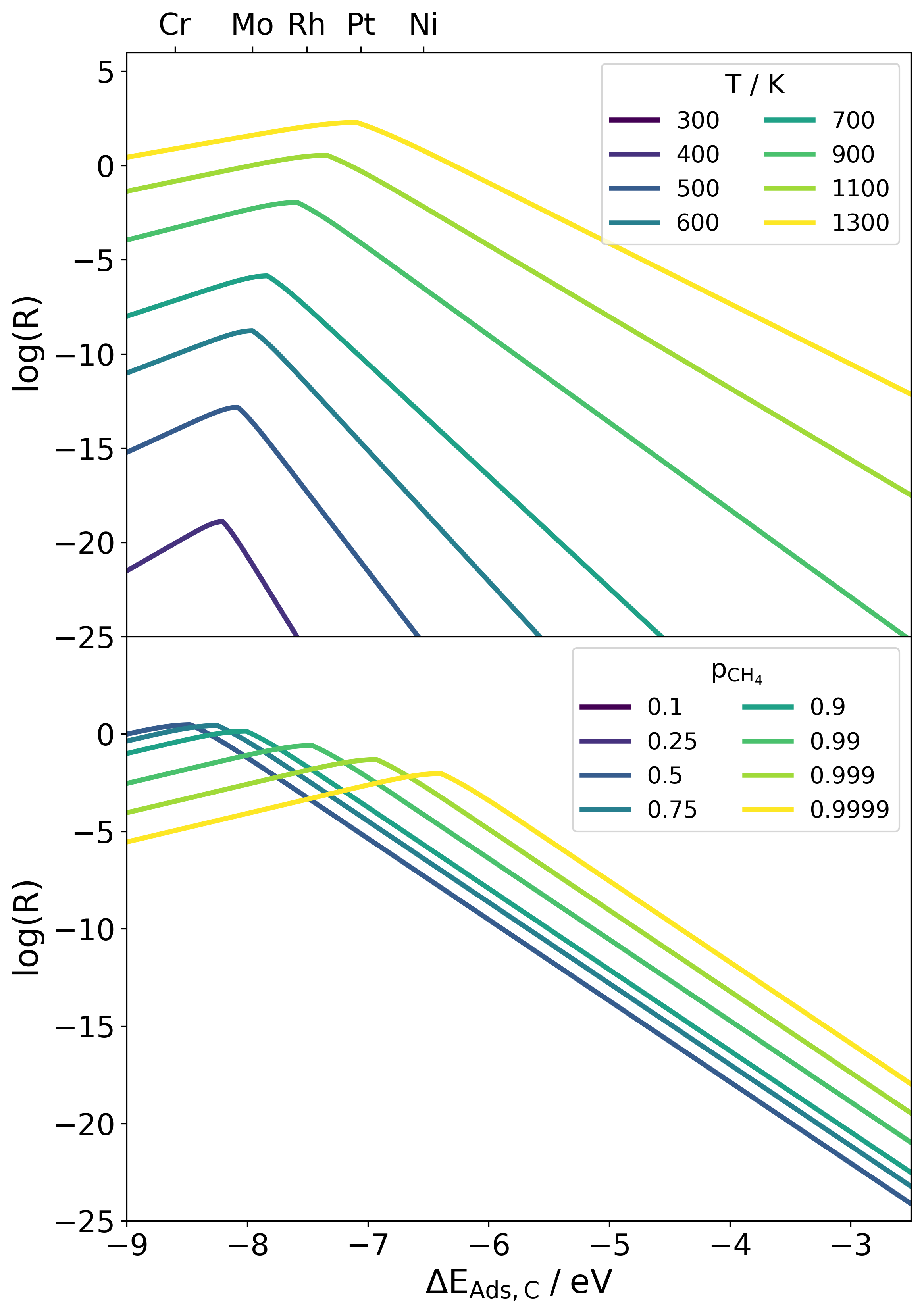}
    \captionsetup{width=.95\linewidth, justification=centering}
    \caption{Top: $\log (R)$ as a function of $\Delta E_\mathrm{Ads, C}$ for different temperatures ($p_{\ce{CH4}} = 0.99$ and $p_{\ce{H2}} = 0.01$). Bottom: $\log(R)$ as a function of $\Delta E_\mathrm{Ads, C}$ for different pressures ($T = 1000$ K). A selection of elements is displayed on the top axis to illustrate their respective positions.}
    \label{R(T, p)}
\end{figure}

\begin{figure}[ht!]
    \centering
    \includegraphics[width=0.99\textwidth]{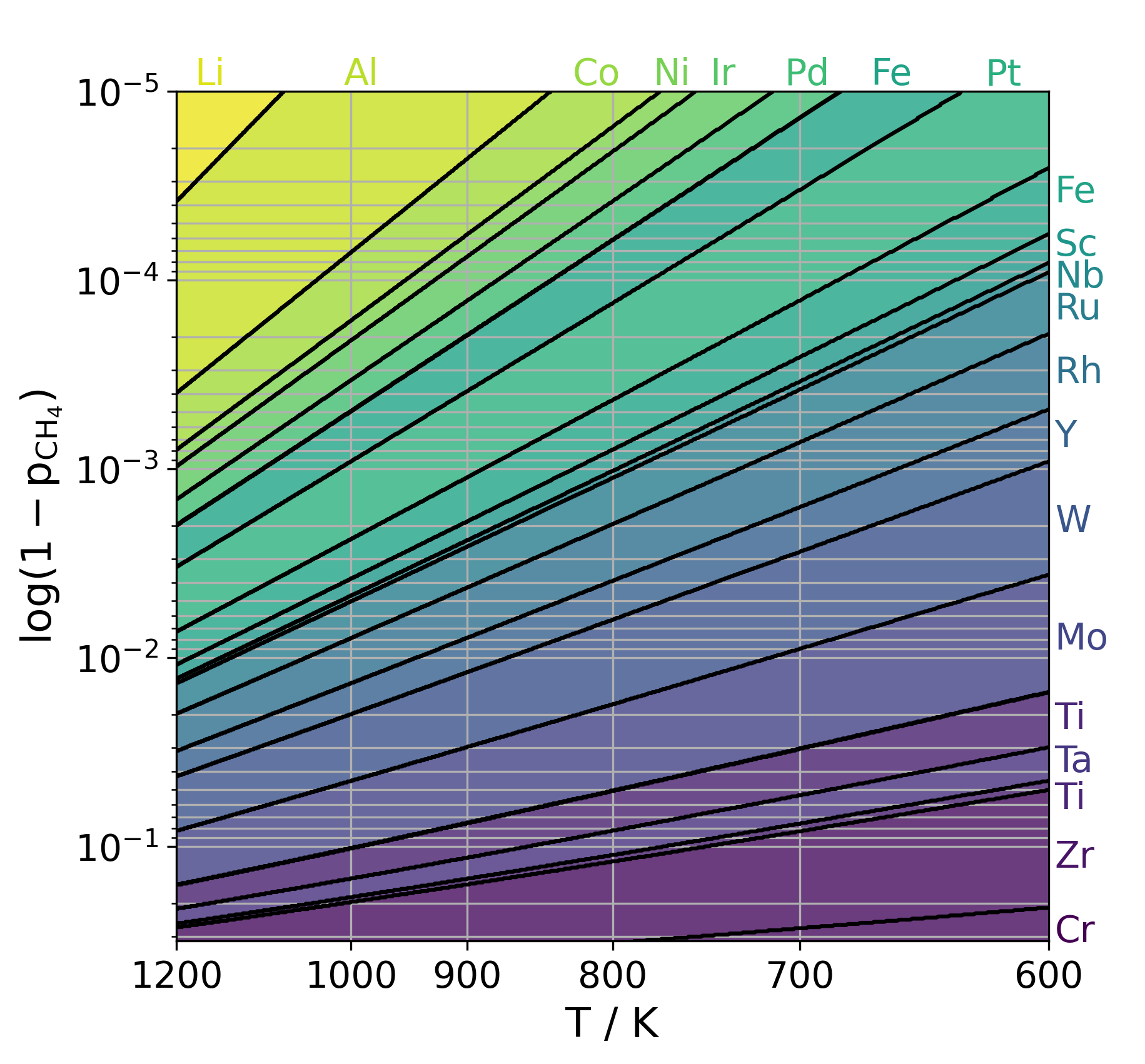}
    \captionsetup{width=.9\linewidth, justification=centering}
    \caption{Arrhenius plot of the ideal catalyst as a function of $1/T$ and $\log(p_{\ce{H2}})= \log(1 - p_{\ce{CH4}})$. Note that the colors have no meaning other than to help distinguish areas with different optimal catalysts.}
    \vspace{0pt}
    \label{cat_diagram}
\end{figure}

In Fig.~\ref{cat_diagram}, the respective best-performing catalyst is represented in an Arrhenius diagram. The corresponding catalysts are listed at the edges of the diagram. It is clearly visible how the optimum moves towards lower magnitudes of adsorption energy with increasing temperature and partial pressure of \ce{CH4}, as already shown in Fig.~\ref{R(T, p)}. In total, 19 different elements are closest to the maximum of $R$ at least once for a temperature range between 600 and 1200 K and a partial pressure range of \ce{H2} between $10^{-1}$ and $10^{-5}$, suggesting that the optimal operating conditions and catalyst material are closely matched. Hence, such a diagram could be useful for practical designing the pyrolysis process.

\subsection{Limitations of the present model}
The presented model efficiently trades off between computational effort and its predictive power. Due to the complex nature of a multi-step catalytic chemical reaction with a variety of possible side reactions, a complete mathematical description of the process is considered virtually impossible. We have therefore limited the operating temperature to 1300 K, up to which methane remains the thermodynamically most stable hydrocarbon \cite{Guret1997}. The most important limitations are imposed by surface geometry and molecule orientation with respect to the surface, imperfect correlation coefficients, neglected consideration of reaction kinetics and carbon, the choice for the rate-determining step as well as computational accuracy. These factors should be considered in further studies. It is also envisioned that the inclusion of modern machine learning algorithms\cite{Ulissi2017} based on the scarce DFT data will provide more realistic and accurate scaling relationships of adsorption energies.

\section{Conclusions}
In this paper, a model to quantify catalyst materials for methane pyrolysis based on their reaction rate was derived and presented. The approach is based on several chemical models, such as the Sabatier, transition-state, and microkinetic model, which were modified for the specifics of the methane pyrolysis reaction. Required input energies were obtained from density functional theory (DFT) calculations. The resulting model computes a reaction rate that is solely a function of temperature $T$, partial pressures of \ce{CH4} and \ce{H2} gasses, and adsorption energy scaling relations, which are functions of the adsorption energy of carbon $\Delta E_\mathrm{Ads, C}$. The model is considered to be valid for any catalyst material that follows linear adsorption energy scaling relationships and yields a qualitative comparison of the materials.

Furthermore, the model enables the investigation of the best-performing catalyst material in the $T$--$p$ space for a given database of adsorption energies on a set of pure fcc, bcc and hcp metals. For a temperature range between 600 and 1200 K and a partial pressure range for \ce{H2} between $10^{-1}$ and $10^{-5}$, a total of 18 different metals were found to be optimal catalyst materials at least once. 

Finally, we point out that our study demonstrates that the type of catalyst and specific reactor operating conditions should be matched.

\begin{acknowledgement}
The computational results were, in part, achieved by using the Vienna Scientific Cluster computing infrastructure.
\end{acknowledgement}

\begin{suppinfo}
Additional details on computational methods are available as a separate document (PDF).
\end{suppinfo}

\bibliography{pyrolysis_model}

\clearpage
\thispagestyle{empty}
\begin{figure}[b!]
    \centering
    \includegraphics[width=8.25cm,height=4.4cm]{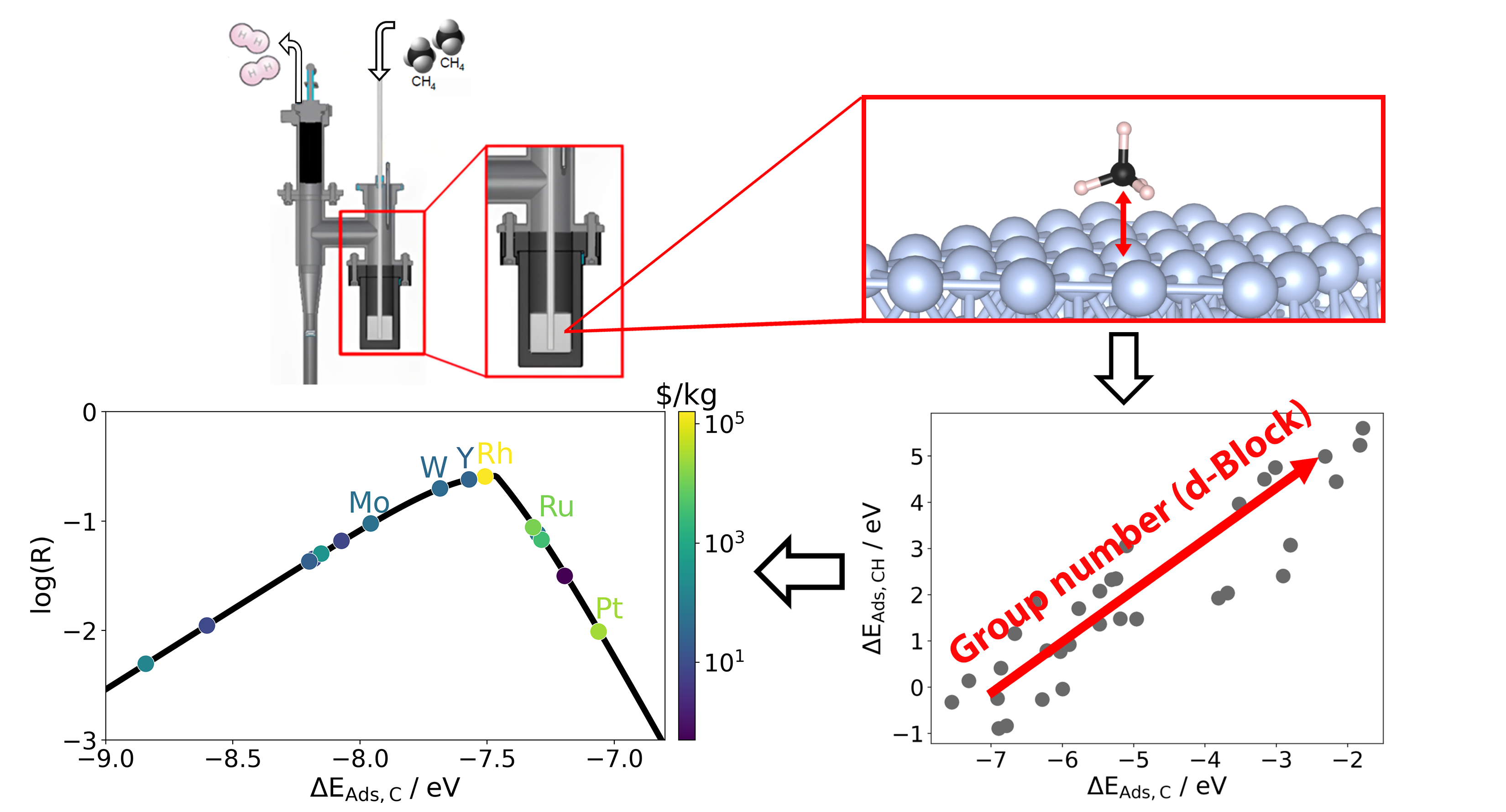}
\end{figure}

\end{document}